\documentclass[aps,showpacs,floatfix,superscriptaddress,nofootinbib]{revtex4}
\usepackage{graphicx}
\usepackage{amsmath}

\newcommand {\be} {\begin{equation}}
\newcommand {\ee} {\end{equation}}
\newcommand {\Be}{\begin{eqnarray*}}
\newcommand {\Ee} {\end{eqnarray*}}
\newcommand {\bey} {\begin{eqnarray}}
\newcommand {\eey} {\end{eqnarray}}
\newcommand{\bit}{\begin{itemize}}      
\newcommand{\eit}{\end{itemize}}
\newcommand{\bfl}{\begin{flusleft}}
\newcommand{\efl}{\end{flusleft}}
\newcommand{\bfr}{\begin{flushright}}
\newcommand{\ec}{\end{center}}
\newcommand{\ben}{\begin{enumerate}}    
\newcommand{\een}{\end{enumerate}}

\newcommand{\comment}[1]{}

\begin{document} 

\title{Average synaptic activity and neural networks topology: a global inverse problem}
\author{Raffaella Burioni}
\affiliation{Dipartimento di Fisica e Scienza della Terra,  Universit\`a di
Parma, via G.P. Usberti, 7/A - 43124, Parma, Italy}
\affiliation{INFN, Gruppo Collegato di Parma, via G.P. Usberti, 7/A - 43124, Parma, Italy} 
\author{Mario Casartelli}
\affiliation{Dipartimento di Fisica e Scienza della Terra,  Universit\`a di
Parma, via G.P. Usberti, 7/A - 43124, Parma, Italy}
\affiliation{INFN, Gruppo Collegato di Parma, via G.P. Usberti, 7/A - 43124, Parma, Italy} 
\author{Matteo di Volo}
\affiliation{Dipartimento di Fisica e Scienza della Terra,  Universit\`a di Parma, via G.P. Usberti, 7/A - 43124, Parma, Italy}
\affiliation{Centro Interdipartimentale per lo Studio delle Dinamiche Complesse, via Sansone, 1 - 50019 Sesto Fiorentino, Italy}
\affiliation{INFN, Gruppo Collegato di Parma, via G.P. Usberti, 7/A - 43124, Parma, Italy}
\author{Roberto Livi}
\affiliation{Dipartimento di Fisica,  Universit\`a di Firenze, via Sansone, 1 - 50019 Sesto Fiorentino, Italy}
\affiliation{Istituto dei Sistemi Complessi, CNR, via Madonna del Piano 10 - 50019 Sesto Fiorentino, Italy}
\affiliation{INFN Sez. Firenze, via Sansone, 1 -50019 Sesto Fiorentino, Italy}
\affiliation{Centro Interdipartimentale per lo Studio delle Dinamiche
Complesse, via Sansone, 1 - 50019 Sesto Fiorentino, Italy}
\author{Alessandro Vezzani}
\affiliation{ S3, CNR Istituto di Nanoscienze, Via Campi, 213A - 41125 Modena, Italy}
\affiliation{Dipartimento di Fisica e Scienza della Terra,  Universit\`a di Parma, via G.P. Usberti, 7/A - 43124, Parma, Italy}
\begin{abstract}
The dynamics of neural networks is often characterized by collective behavior and quasi-synchronous events,
where a large fraction of neurons fire in short time intervals, separated by uncorrelated firing activity.  These global temporal signals are crucial for brain functioning. They strongly depend on the topology of the network and on the fluctuations of the connectivity. We propose a {\sl heterogeneous mean--field} approach to neural dynamics on random networks, that explicitly preserves the disorder in the topology at growing network sizes, and leads to a set of self-consistent equations. Within this approach, we provide an effective description of microscopic and large scale temporal signals in a leaky integrate-and-fire model 
with short term plasticity, where quasi-synchronous events arise. Our equations provide a clear analytical picture of the dynamics, evidencing the contributions of both periodic (locked) and aperiodic (unlocked) neurons to the measurable average signal.  In particular, we formulate and
solve a {\sl global} inverse problem of reconstructing the in-degree
distribution from the knowledge of the average activity field.  Our method is very general and applies to a large class of dynamical models on dense random networks. 
\end{abstract}
   

\maketitle


\section*{Introduction}

Topology has a strong influence on phases of dynamical models defined on a network. Recently, 
this topic has attracted the interest of both theoreticians
and applied scientists in many
different fields, ranging from physics, to biology and social sciences \cite{Xiao,vespignani,mendes,cohenhavlin,Arenas,Boccaletti}. Research has focused
in two main directions. The {\sl direct problem} aims at
predicting the dynamical properties of a network from its
topological parameters \cite{donetti}. The {\sl inverse problem}
is addressed to the reconstruction of the network topological features
from dynamic time series \cite{inverse1,inverse1A,inverse2,aurell}. 
The latter approach is particularly interesting when the 
direct investigation of the network is impossible or very hard 
to be performed. 

Neural networks are typical examples of such a situation. In {\sl local} approaches to inverse problems \cite{inverse1,inverse1A,inverse2,aurell}, the network is reconstructed through the knowledge of long time series of single neuron dynamics, a methods that applies efficiently to small systems only.  Actually, the signals emerging during neural time evolution
are often records of the average synaptic activity from large regions 
of the cerebral cortex -- 
a kind of observable likely much easier to be measured
than signals coming from single neuron
activities \cite{fmri,fmri1}. Inferring the topological properties of the network
from global signals is still an open and central
problem in neurophysiology. In this paper we investigate the possibility
of formulating and solving such a {\sl global} version of the inverse
problem, reconstructing the network topology that has generated a given global (i.e. average)
synaptic-activity field. The solution of such an inverse problem
could also imply the possibility of engineering a network able to
produce a specific average signal. 

As an example of neural network dynamics, we focus on a system of leaky
integrate--and--fire (LIF) excitatory neurons, interacting
via a synaptic current regulated by the short--term 
plasticity  mechanism \cite{plast1,plast2}. This model is able to reproduce synchronization
patterns observed in {\sl in vitro} experiments
\cite{tsodyksnet,volman,DL}. As a model for the underlying topology we consider randomly uncorrelated
diluted networks made of $N$ nodes. In general $N$ is 
considered quite a 
large number, as is the number of connections between pairs
of neurons. This suggests that the right framework for understanding
large--population neural networks should be a mean--field approach,
where the thermodynamic limit, $N \to \infty$,  is expected to provide
the basic ingredients for an analytic treatment. On the other hand, 
the way such a thermodynamic limit is performed may
wipe out any relation with the topological features that are responsible,
for finite $N$, of relevant dynamical properties.

In  Erd\"os--Renyi directed networks, where each neuron is
randomly and uniformly connected to a finite fraction of
the other neurons ({\sl massive or dense connectivity}), the
fluctuations of the degree determine a rich dynamical behavior,
characterized in particular  by {\sl quasi-synchronous events} (QSE). This
means that a large fraction of neurons fire in a short time interval of
a few milliseconds (ms), separated by uncorrelated firing activity
lasting over some tens of ms. Such an interesting behavior
is lost in the thermodynamic limit, as the fluctuations
of the connectivity vanish and the "mean-field-like" dynamics
reduces to a state of fully synchronized neurons (e.g., see \cite{DLLPT}).
In order to maintain the QSE phenomenology  in the large $N$ limit, we
can rather consider the sequence of random graphs that keep the same 
specific in-degree distribution $P(\tilde k)$, where ${\tilde k}_i = k_i/N$ is the fraction of incoming neurons
connected to neuron $i$ for any finite $N$,
similarly to the configuration models \cite{confi}.
This way of performing the thermodynamic limit preserves the dynamical regime of QSE and the
difference between synchronous and non-synchronous
neurons according to their specific in-degree $\tilde k$. By introducing explicitly
this $N \to \infty$ limit in the differential equations of
the model, we obtain a {\sl heterogeneous mean--field} (HMF)
description, similar to the one recently introduced
in the context of epidemiological spreading on networks \cite{vespignani,mendes,vesp1,vesp2}.
Such mean--field like  or HMF equations
can be studied analytically by introducing the return maps of the firing times. 
In particular, we find that a sufficiently
narrow distributions of $P(\tilde k)$ is necessary to observe
the quasi--synchronous dynamical phase, which vanishes on the contrary for broad
distributions of $\tilde k$.

More importantly, these HMF equations allow us to
design a "global" inverse--problem approach, formulated 
in terms of an integral Fredholm equation of the first kind 
for the unknown  $P(\tilde k)$ \cite{kress}.
Starting  from the dynamical
signal of the average synaptic-activity field, the solution of this equation provides with good
accuracy the $P(\tilde k)$ of the network that produced it. We test
this method for very different uncorrelated network topologies, where $P(\tilde k)$ ranges
from a Gaussian with one or several peaks, to power law
distributions, showing its effectiveness even for finite size networks.

The overall procedure
applies to a wide class of network dynamics of the type
\begin{equation}
\label{eq1n}
 \dot{\textit{\textbf{w}}_i} =  \mathbf{F} \Big ({\textit{\textbf{w}}}_i,\frac{g}{N}\sum_{j \ne i} \epsilon_{ji} G(\textit{\textbf{w}}_j)\Big )  \,\,\, ,  \,
\end{equation}
where the vector $\textit{\textbf{w}}_i$ represents the state of the site $i$, $ \mathbf{F}(\textit{\textbf{w}}_i,0) $ is the single site dynamics, $g$ is the coupling strength, $G(\textit{\textbf{w}}_j)$ is a suitable coupling function  and $\epsilon_{j,i}$ is the  adjacency matrix of the directed uncorrelated dense network, whose entries are equal to $1$ if  neuron $j$ fires to neuron $i$, and $0$ otherwise.

\section*{Results}
\subsection*{The LIF model with short term plasticity}

Let us introduce LIF models, that describe a network  of $N$ neurons
interacting via a synaptic current, regulated by short--term--plasticity with equivalent synapses \cite{tsodyksnet}. In this case
the dynamical variable of the neuron $i$ is 
$ \textit{\textbf{w}}_i= (v_i,x_i,y_i,z_i)$  where $v_i$ is the rescaled membrane potential and  $x_i$, $y_i$, and $z_i$ represent the
fractions of synaptic transmitters in the recovered, active, and inactive state, 
respectively ($x_i+y_i+z_i=1$).  Eq. (\ref{eq1n}) then specializes to:

\begin{align}
\label{dynv}
& \dot v_i= a -v_i + \frac{g}{N} \sum_{j \ne i} \epsilon_{ji} y_j \\
\label{dynsyn}
& \dot y_{i} = -\frac{y_{i}}{\tau_{\mathrm{in}}} +ux_{i}S_i\\
\label{contz}
& \dot z_{i} = \frac{y_{i}}{\tau_{\mathrm{in}}}  -   \frac{z_{i}}{\tau_{\mathrm{r}}} \,\,\, .
\end{align} 
The function  $S_j(t)$ is the spike--train produced by neuron $j$, $S_j(t)=\sum_m \delta(t-t_{j}(m))$, where $t_{j}(m)$ is the time when neuron $j$ fires its $m$-th spike. Notice that we assume the spike to be a $\delta$ function of time. Whenever the potential $v_i(t)$ crosses the threshold value $v_{\mathrm{th}}=1$, it is 
reset to $v_{\mathrm{r}}=0 $, and a spike is sent towards its efferent neurons. 
The mechanism of short--term plasticity, that mediates the transmission of the field $S_j(t)$,
was introduced in \cite{plast1,plast2} to account for the activation of neurotransmitters
in neural dynamics mediated by synaptic connections. 
In particular, when neuron $i$ emits a spike, it releases a fraction of neurotransmitters  $u x_i(t)$  (see the second term
in the r.h.s. of Eq. (\ref{dynsyn}) ), and the  fraction of active resources $y_i(t)$ is increased. Between consecutive spikes of neuron $i$, the use of active resources determines the exponential decrease of $y_i(t)$, on a time scale $\tau_{\mathrm{in}}$, thus yielding the increment of the fraction  of inactive resources $z_i(t)$ (see the first term on the r.h.s. of Eq. (\ref{contz})). 
Simultaneously, while  $z_i(t)$ decreases (see the second term on the r.h.s. of Eq. (\ref{contz})), the fraction of available resources is  recovered over a time scale $\tau_{\mathrm{r}}$:
in fact, from Eq.s (\ref{dynsyn}) and (\ref{contz}) one readily obtains $\dot x_i(t)=z_i(t)/\tau_{\mathrm{r}} -ux_i(t)S_i(t)$.
We assume that all parameters appearing in the above 
equations are independent of the neuron indices, and that each neuron
is connected to a macroscopic number, 
${\mathcal O}(N)$, of pre-synaptic neurons: this is the reason why the interaction term is divided by the 
factor $N$.  In all data hereafter reported we have used phenomenological values of the
rescaled parameters:
$\tau_\mathrm{in} = 0.2$,  $\tau_{\mathrm{r}} = 133\tau_{\mathrm{in}}$, 
$a= 1.3$,  $g = 30$ and $u = 0.5$ \cite{DLLPT}. 

The choice of the value of the external current, $a$, is quite important for selecting the dynamical regime
one is interested to reproduce. In fact, for $a>v_{th}=1$,  neurons 
are in a firing regime, that tyically gives rise to collective 
oscillations \cite{DLLPT,olmi, luccioli,brunel}.  These have been observed experimentally 
in mammalian brains, where such a coherent rythmic behavior involves 
different groups of neurons  \cite{busaki}.  On the other hand, it is also well 
known that in many cases neurons operate in the presence of a subthreshold 
external current \cite{DL}. In this paper, we aim to present an approach that
works irrespectively of the choice of $a$. For the sake of simplicity,
we have decided to describe it for $a=1.3$, i.e. in a strong firing regime.

Numerical simulations can be performed effectively by transforming 
the set of differential equations (\ref{dynv})--(\ref{contz}) into an event--driven map \cite{DLLPT,brette,zill}. 
On Erd\"os--Renyi random graphs, where each link is accepted with probability $p$, so that the average in-degree $\langle k \rangle = pN$, the dynamics has been analyzed in detail \cite{DLLPT}. Neurons separate spontaneously into two different families: the {\it locked}  and the {\sl unlocked} ones.  The locked neurons determine quasi-synchronous events (QSE) 
and exhibit a periodic dynamics. The unlocked ones participate in the uncorrelated firing activity and exhibit a sort of irregular evolution. Neurons belong to one of the two families according to their in-degree $k_i$.
 In this topology, the thermodynamic limit can be simply worked out. Unfortunately,  this misses all the interesting features emerging from the model at finite $N$.
Actually, for any finite value of $N$, the in--degree distribution $P(k)$ is centered around $\langle k \rangle$, with a standard deviation $\sigma_k \sim  N^{\frac{1}{2}}$. 
The effect of disorder  is quantified by the ratio $\sigma_k /\langle k \rangle$, that vanishes for $N\to \infty$. Hence  the thermodynamic limit reproduces the naive mean--field like  dynamics of  
a fully coupled network, with rescaled coupling $g\to pg$, that is known to  eventually  converge to a periodic fully synchronous state \cite{DLLPT}.

\subsection*{The LIF model on random graphs with fixed in--degree distribution}

\begin{figure}[t!]
\centering
\includegraphics[scale=0.35]{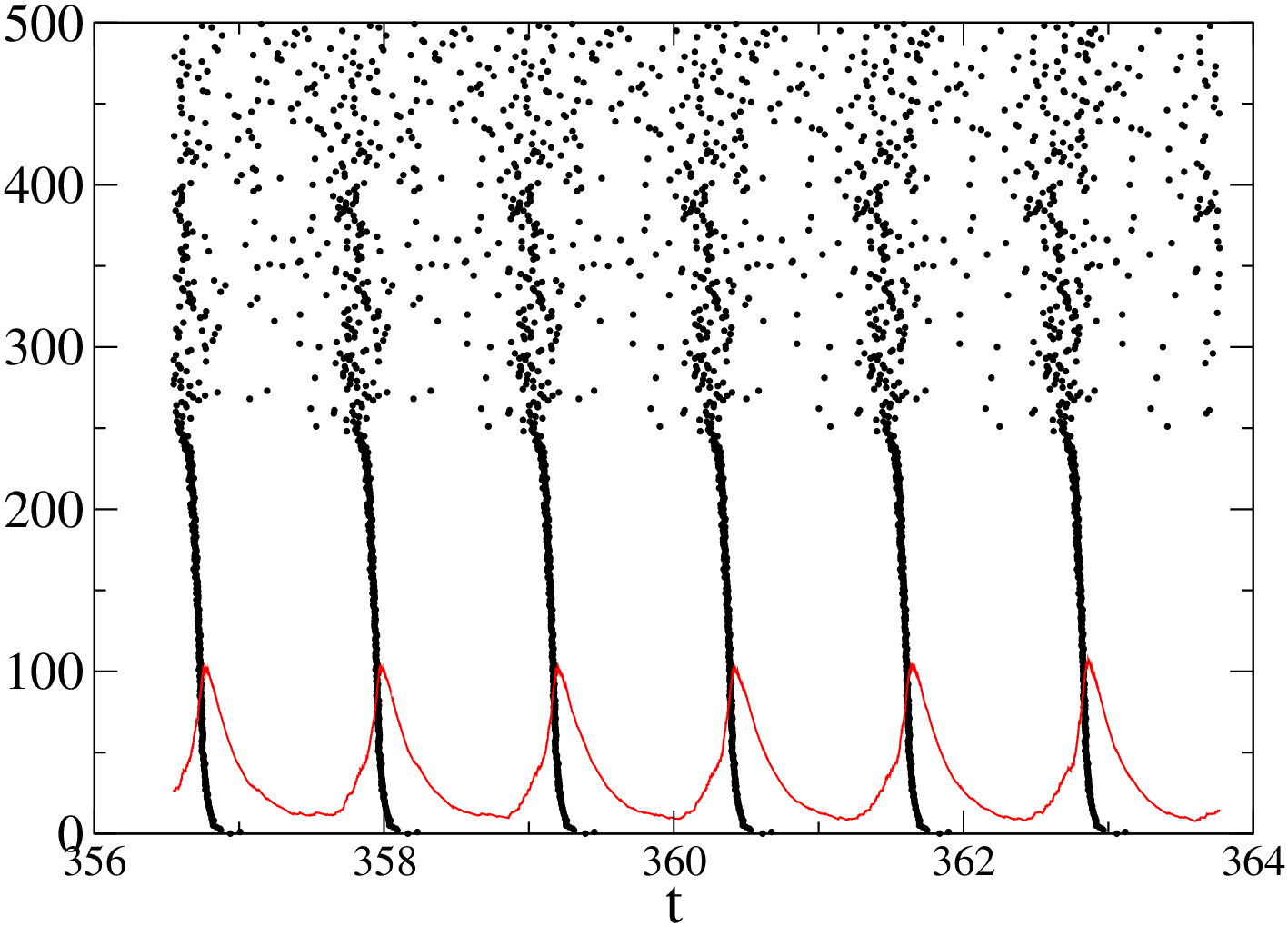}
\caption{Raster plot representation of the dynamics of a network of $N=500$ neurons with a Gaussian in-degree distribution $P(\tilde k)$
($\langle \tilde k \rangle=0.7$,  $\sigma_{\tilde k}=0.077$). Black dots signal single firing events of neurons at time $t$.
Neurons are naturally ordered along the vertical axis according to the values of their in-degree. 
The  global field $Y(t)$ (red line) is superposed to the raster plot for comparison (its actual
values are multiplied by the factor $10^3$, to make it visible on the vertical
scale). }
\label{rp1}
\end{figure}

\begin{figure}[t!]
\centering
\includegraphics[scale=0.35]{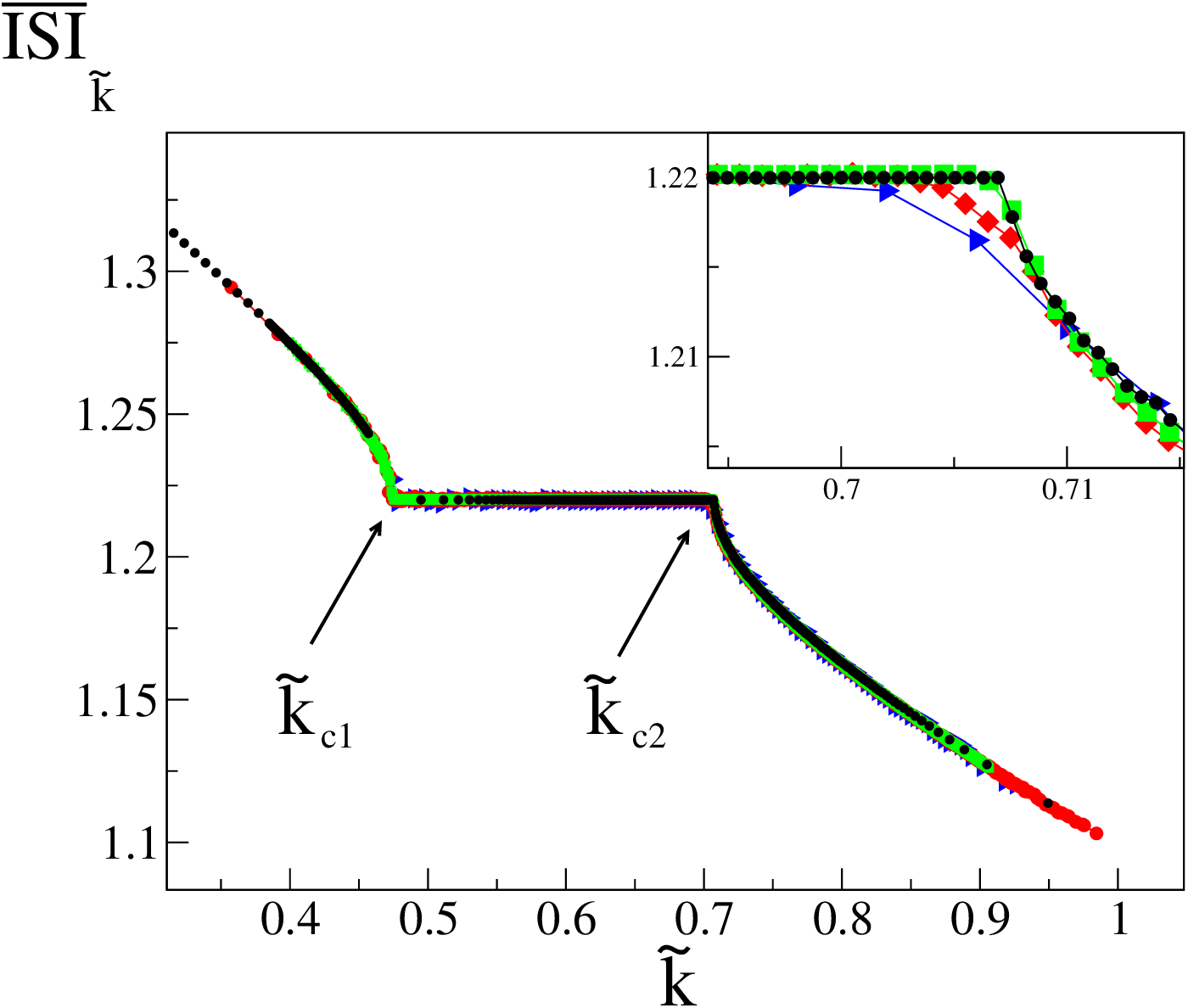}
\caption{Time average of inter-spike intervals $\overline{ISI}({\tilde k})$ vs $\tilde k$ from a Gaussian distribution  with  $\langle \tilde k \rangle=0.7$ and $\sigma_{\tilde k}=0.077$
and for three networks with $N=500$ (blue triangles), $N=5000$ (red diamonds), $N=20000$ (green squares). For each size, the average is taken over 8 different realizations of the random network.  We have also performed a suitable binning over the values of $\tilde k$, thus yielding the numerical 
estimates of the critical values $\tilde k_{c1} \approx 0.49$ and $\tilde k_{c2} \approx 0.70$.
 In the inset we show a zoom of the crossover region  close to $\langle \tilde k \rangle =0.7$. Black dots are the result of simulations of the mean field dynamics
(see Eq.s  (6) -- (8)) with $M=307$. }
\label{2}
\end{figure}

At variance with the network construction discussed in previous sections, uncorrelated random graphs can be defined by different protocols,  that keep track of the in-degree inhomogeneity
 in the thermodynamic limit. In our construction, we fix the
normalized in--degree probability distribution $P(\tilde k)$, so that
$\sigma_k /\langle k \rangle$ is kept constant for any $N$ \cite{confi}. 
Accordingly, $P(\tilde k)$ is a normalized distribution defined in the
interval $\tilde k \in (0,1]$ (while the number of inputs $k\in(0,N]$). 
In particular, if  $P(\tilde k)$ is a truncated Gaussian distribution, the dynamics reproduces the scenario observed in
\cite{DLLPT} for an Erd\"os--Renyi random graph.  
In fact, also in this case neurons are dynamically distinguished into two families,  depending on their in--degree. 
Precisely, neurons with  $\tilde k$ in between two {\sl critical values},  $\tilde k_{c1}$ and $\tilde k_{c2} \approx \langle \tilde k \rangle$,
are locked and determine the QSE: they fire with almost the same period, but exhibit different ($\tilde k$-dependent) phases.
All the other neurons are unlocked and fire in between QSE displaying an aperiodic behavior. Notice that
the range $0 < \tilde k < \tilde k_{c1}$ corresponds to the left tail of the truncated Gaussian distribution; accordingly,
the large majority of unlocked neurons is found in the range  $\tilde
k_{c2} < \tilde k < 1$ (see Fig. \ref{rp1}).
 In order to characterize the dynamics at increasing  $N$, we consider for each neuron its 
inter-spike-interval (ISI), i.e. the lapse of time in between
consecutive firing events. In Fig.\ref{2}  we show  the time-average
of ISI vs  $\tilde k$, or $\overline{ISI}(\tilde k)$. One can clearly observe the plateau of locked
neurons and the crossover to unlocked neurons at the critical values $\tilde k_{c1}$ and
$\tilde k_{c2}$. 
 
Remarkably, networks of different sizes ($N=500,~5000$ and $20000$)
feature the same  $\overline{ISI}(\tilde k)$ for locked neurons, and almost the same  
values of $\tilde
k_{c1}$ and $\tilde k_{c2}$. 
There is not a sharp transition from locked to unlocked neurons,  
because for finite $N$  the behavior of each neuron
depends not only on its  $\tilde k$, but also on neighbor neurons sending their inputs. 
Nevertheless, in the inset, the crossover appears to be sharper and sharper for increasing $N$, as expected for true critical points. Furthermore, the fluctuations of 
$\overline{ISI}(\tilde k)$
over different realizations, by $P(\tilde k)$,  of  three networks of different size exhibit a peak around  $\tilde k_{c1}$ and $\tilde k_{c2}$, while they decrease with $N$ as 
$\sim N^{-1/2}$ (data not shown). Thus, the qualitative and quantitative features of the QSE at finite sizes are expected to persist in the thermodynamic limit, where
fluctuations vanish and  the dynamics of each neuron depends only  on its in--degree.

\subsection*{Heterogeneous Mean Field equations}

We can now construct the Heterogeneous Mean--Field (HMF) equations for our model
by combining this thermodynamic limit procedure with a  consistency relation in Eqs. (\ref{dynv}) - (\ref{contz}).
The input field received by neuron $i$ is $Y_i= 1/N\sum_j\epsilon_{ij}y_j(t) =
1/N\sum_{j\in I(i)} y_j$, where $I(i)$ is the set of $k_i$ neurons transmitting to
neuron $i$. For very large values of $N$ the average field generated by presynaptic neurons of neuron
$i$, i.e. $1/k_i \sum_{j\in I(i)} y_j$, can be approximated with $1/N\sum_j y_j$, where the
second sum runs over all neurons in the network ({\sl mean--field hypothesis}).
Accordingly, we have $Y_i=(k_i/N) (1/k_i)\sum_{j\in I(i)} y_j \approx \tilde k_i (1/N\sum_j y_j)$: as a consequence in the thermodynamic limit the dynamics of each neuron depends
only on its specific in--degree. In this limit,
 $\tilde k_i$ becomes a continuous variable   independent of the label $i$, taking values in the
 interval (0,1].  Therefore, all neurons with the same in--degree $\tilde k$ follow the same dynamical equations and we can write the dynamical equations for the class of neurons with in--degree $\tilde k$.
Finally, we can  replace $Y_i $ with $\tilde k Y(t)$, where  
\begin{equation}
\label{meanfield}
Y(t)=\int_{0}^{1}P(\tilde k) y_{\tilde k}(t)d\tilde k
\end{equation}
The HMF equations, then, read
\begin{align}
\label{vk}
&\dot v_{\tilde k}(t)= a -v_{\tilde k}(t) + g\tilde kY(t)\\
\label{yk}
& \dot y_{\tilde k}(t) = -\frac{y_{\tilde k}(t)}{\tau_{\mathrm{in}}} +u(1-y_{\tilde k}(t)-z_{\tilde k}(t))S_{\tilde k}(t)\\
\label{zk}
& \dot z_{\tilde k}(t) = \frac{y_{\tilde k}(t)}{\tau_{\mathrm{in}}}  -   \frac{z_{\tilde k}(t)}{\tau_{\mathrm{r}}} \,\,\, ,
\end{align}
where $v_{\tilde k}$, $y_{\tilde k}$ and $z_{\tilde k}$ are the membrane potential, fraction of active and inactive resources of the class of neurons with in--degree $\tilde k$, respectively.
Despite this set of equations cannot be solved explicitly,  they provide a great numerical advantage with respect to direct simulations
of large systems. Actually, the basic features of the  dynamics of such systems can be effectively reproduced (modulo finite--size corrections) by 
exploiting a suitable sampling of $P(\tilde k)$. For instance,  one can sample the continuous variable $\tilde k \in [0,1] $ into $M$ discrete
values $\tilde k_i$ in such a way that  $\int_{\tilde k_i}^{\tilde k_{i+1}}P(\tilde k)d\tilde k$ is kept fixed (importance sampling).    
Simulations of Equations (\ref{meanfield})-(\ref{zk}) show that 
the global field field $Y(t)$ is periodic and the neurons split into locked and unlocked. 
Locked neurons feature a firing time delay with respect the peak of $Y(t)$, and this phase shift
depends on the in--degree $\tilde k$. As to unlocked neurons, that typically 
have an in-degree $\tilde k>\langle \tilde  k \rangle$, they follow a complex dynamics with irregular firing times. 
In Fig. \ref{2} (black dots) we compare $\overline {ISI} (\tilde k)$,  obtained from the HMF equations for
$M= 307$, with the same quantity computed by direct simulations of networks up to size $N=2 \times 10^4$. 
The agreement is remarkable evidencing the numerical effectiveness of the method. 

\subsection*{The direct problem: stability analysis and the synchronization transition}

In the HMF equations, once the global field $Y(t)$ is known, the dynamics of each class of neurons 
with in-degree $\tilde k$ can be determined by a straightforward integration, and we can
perform the stability analysis that Tsodyks et al. applied to a similar model \cite{tso_locked}.  As an example, 
we have considered the system studied in Fig.\ref{2}. The global field $Y(t)$ of the HMF dynamics has been obtained using 
the importance sampling for the distribution 
$P(\tilde k)$.  We have fitted $Y(t)$  with an analytic function of time $Y_f(t)$, continuous and periodic in time, with period $T$. 
Accordingly,  Eq. (6) can be  approximated by
\begin{equation}
\label{s1}
\dot v_{\tilde k}(t)= a -v_{\tilde k}(t) + g\tilde kY_f(t).
\end{equation}
\vskip 30pt

\begin{figure}[t!]
\centering
\includegraphics[scale=0.35]{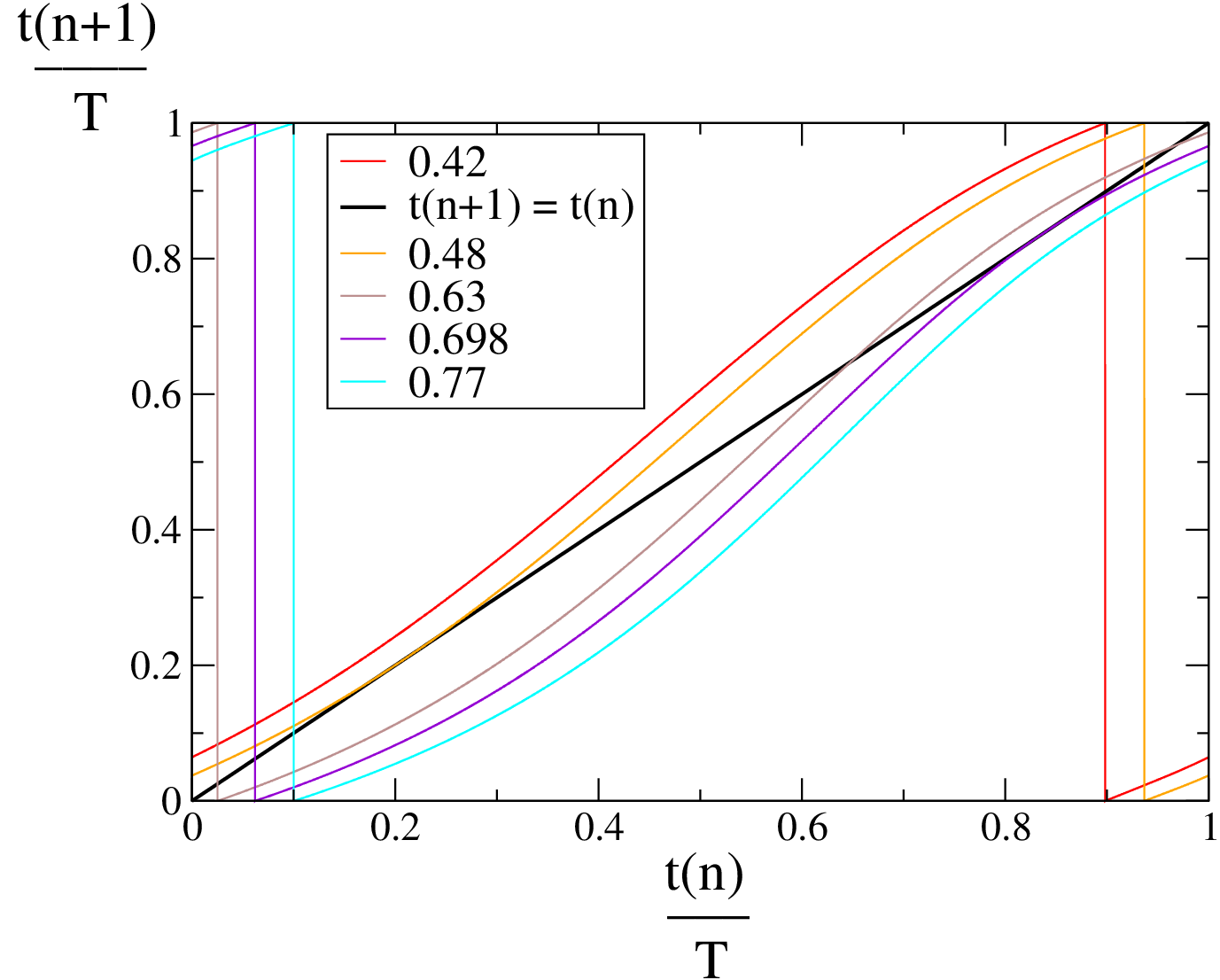}
\caption{The return map $R_{\tilde k}$ in Eq. (10) of the rescaled variables $t_{\tilde k}(n)/T$ for different values of $\tilde k$, corresponding to lines of different colors, according to the legend in the inset: the black line is the bisector of the square.}
\label{6}
\end{figure}  
Notice that, by construction, the field $Y_f(t)$ features peaks at times $t=nT$, where $n$ is an integer. In this way
we can represent  Eq. (\ref{s1}) as a discrete single neuron map. In practice, we integrate Eq.(\ref{s1}) and determine the
sequence of the (absolute value of the) firing time--delay, $t_{\tilde k}(n)$, of neurons with in--degree $\tilde k$ with respect to the
reference time $nT$.  The return map $R_{\tilde k}$ of this quantity reads
\begin{equation}
\label{s2}
t_{\tilde k}(n+1)=R_{\tilde k} t_{\tilde k}(n).
\end{equation}
In Fig. \ref{6} we plot the return map of the rescaled firing time--delay  $t_{\tilde k}(n)/T$ for different values of $\tilde k$. We observe that in-degrees $\tilde k$ corresponding to locked neurons (e.g., the brown curve) have two fixed points $t_{\tilde k}^s$ and $t_{\tilde k}^u$, i.e. two points of intersection of the curve with the
diagonal. The first one is stable (the derivative of the map $R_{\tilde k}$ is < 1 ) and the second  unstable (the derivative of the map $R_{\tilde k}$ is > 1). Clearly, the dynamics converges to the stable fixed point displaying a periodic behavior. In particular, the firing times of neurons $\tilde k$ are phase shifted of a quantity $t_{\tilde k}^s$ with respect the peaks of the fitted global field. 
The orange and violet curves correspond to the dynamics at the critical in-degrees $\tilde k_{c1}$ and $\tilde k_{c2}$ where the fixed points disappear (see Fig.(\ref{2})).
The presence of such fixed  points influences also the behavior of the unlocked component (e.g., the red and light blue curves). In particular,
the nearer  $\tilde k$  is to $\tilde k_{c1}$ or to $\tilde k_{c2}$, the closer is the return map to the bisector of the square, giving rise to a dynamics spending longer and longer  times in an almost periodic firing. 
Afterwards, unlocked neurons depart from this almost periodic regime, thus following an aperiodic behavior.
As a byproduct, this dynamical analysis allows to estimate the values of the critical in--degrees. For the system of Fig.\ref{rp1}$, \tilde k_{c1}=0.48$ and $\tilde k_{c2}=0.698$, in very good agreement with the numerical simulations (see Fig. \ref{2}).

\begin{figure}[t!]
\centering
\includegraphics[scale=0.35]{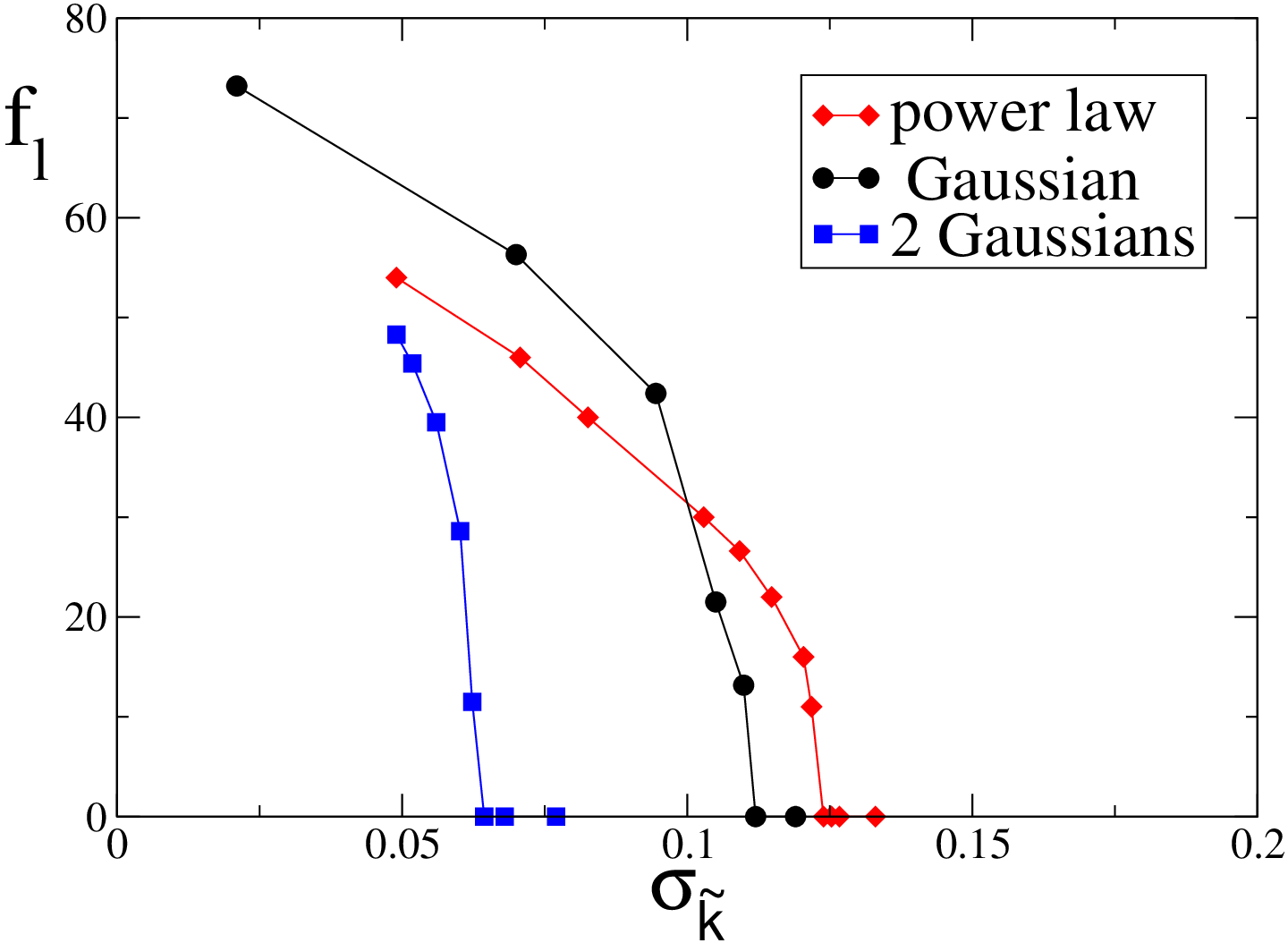}
\caption{The fraction of locked neurons, $f_l$, as a function of the standard deviation $\sigma_{\tilde k}$ of the distributions: truncated 
Gaussian  with $\langle \tilde k \rangle  = 0.7$ (black dots);  truncated superposition of 
two Gaussians (both with standard deviation 0.03), one centered at  $\tilde k_1=0.5$ and the other one at a varying value $\tilde k_2$, that determines the
overall standard deviation $\sigma_{\tilde k}$ (blue squares); truncated power law distribution  with $\tilde k_{\mathrm{min}}=0.1$
(red diamonds). In the last case the value of the standard deviation is changed by varying the exponent $\alpha$, while the average $\langle \tilde k\rangle$ changes accordingly.  Lines have been drawn to guide the eyes.}\label{cross}
\end{figure}

Still in the perspective of the {\sl direct problem}, the HMF equations provide further insight on how the network topology influences the dynamical behavior. We have found that, in general,  the fraction of locked neurons increases as $P(\tilde k)$ becomes sharper and sharper, while synchronization is eventually lost for broader distributions. 
In Fig. \ref{cross} we report the fraction of locked neurons, $f_l=\int_{\tilde k_{c1}}^{\tilde k_{c2}}P(\tilde k)d\tilde k$, as a function of the standard deviation deviation $\sigma_{\tilde k}$, for different kinds of $P(\tilde k)$ (single or double--peaked Gaussian, power law) in the HMF equations.  For the  truncated power law distribution, we set $P(\tilde k)\sim \theta(\tilde k-\tilde k_{\mathrm{min}})\tilde k^{-\alpha}$. In all cases,  there is a critical value of $\sigma_{\tilde k}$ above which $f_l$ vanishes, i.e. QSE disappear. The generality of this scenario
points out  the importance of the relation between $P(\tilde k)$ and the average synaptic field $Y(t)$.
 
\subsection*{The inverse problem}

The HMF approach allows to implement the inverse problem and leads to the reconstruction of the distribution $P(\tilde k)$ from the knowledge of  $Y(t)$. If the global synaptic activity field $Y(t)$ is known, each class of neurons of in-degree $\tilde k$ evolves according to the equations:
\begin{align}
\label{vktil}
&\dot {\mathsf{ v}}_{ \tilde k}(t)= a -\mathsf{ v}_{ \tilde k}(t) + g \tilde kY(t)\\
\label{yktil}
& \dot {\mathsf{ y}}_{ \tilde k}(t) = -\frac{\mathsf{y}_{\tilde k}(t)}{\tau_{\mathsf{in}}} +u(1-\mathsf{ y}_{ \tilde{k}}(t)-\mathsf{ z}_{ \tilde{k}}(t))\tilde S_{ \tilde{k}}(t)\\
\label{zktil}
& \dot {\mathsf{ z}}_{ \tilde k}(t) = \frac{\mathsf{ y}_{ \tilde k}(t)}{\tau_{\mathsf{in}}}  -   \frac{\mathsf{ z}_{ \tilde{k}}(t)}{\tau_{\mathsf{r}}} \,\,\, .
\end{align}
Notice that the variable $\mathsf{v}(t)$, $\mathsf{y}(t)$, $\mathsf{z}(t)$ can take  values that differ from the variables generating the field $Y(t)$, i.e. $v(t)$, $y(t)$, $z(t)$, as they start from different initial conditions.  However,  the self consistent relation for the global field $Y(t)$ implies:
\begin{equation}\label{global}
Y(t)=\int_0^1 P({\tilde k}) \mathsf{y}_{\tilde k}(t)d{\tilde k} \,\,\, .
\end{equation}
If $Y(t)$ and $\mathsf{y}_{\tilde k}(t)$ are known, this is a Fredholm
equation of the first kind in $P(\tilde k)$ \cite{kress}. In the general case of Eq. (\ref{eq1n}), calling $E(t)$ the global measured external field, the evolution equations corresponding to Eq.s (\ref{vktil})--(\ref{zktil}) read
\begin{equation}
{ \dot{\mathbf{w}}_{\tilde k}} =   \mathbf{F} \Big ({\mathbf{w}}_{\tilde k},g \tilde k E(t)\Big ) \,
\end{equation}
and the Fredholm equation for the inverse problem is
\begin{equation}
E(t)=\int_0^1 P(\tilde k) G( {\mathbf{w}}_{\tilde k}(t))d\tilde k \,\,\, .
\end{equation}
In the case of our LIF model, as soon as a locked component exists, Eq. (\ref{global}) can be solved by a functional Montecarlo minimization procedure  applied to a
sampled $P(\tilde k)$. At variance with the {\sl direct problem},  $P(\tilde k)$ is the unknown function and, accordingly, we have to
adopt a  uniform sampling of the support of $\tilde k$.  A sufficiently fine sampling has to be used for
a confident reconstruction of $P(\tilde k)$ (See Methods section).

 \begin{figure}[t!]
\centering
\includegraphics[scale=0.35]{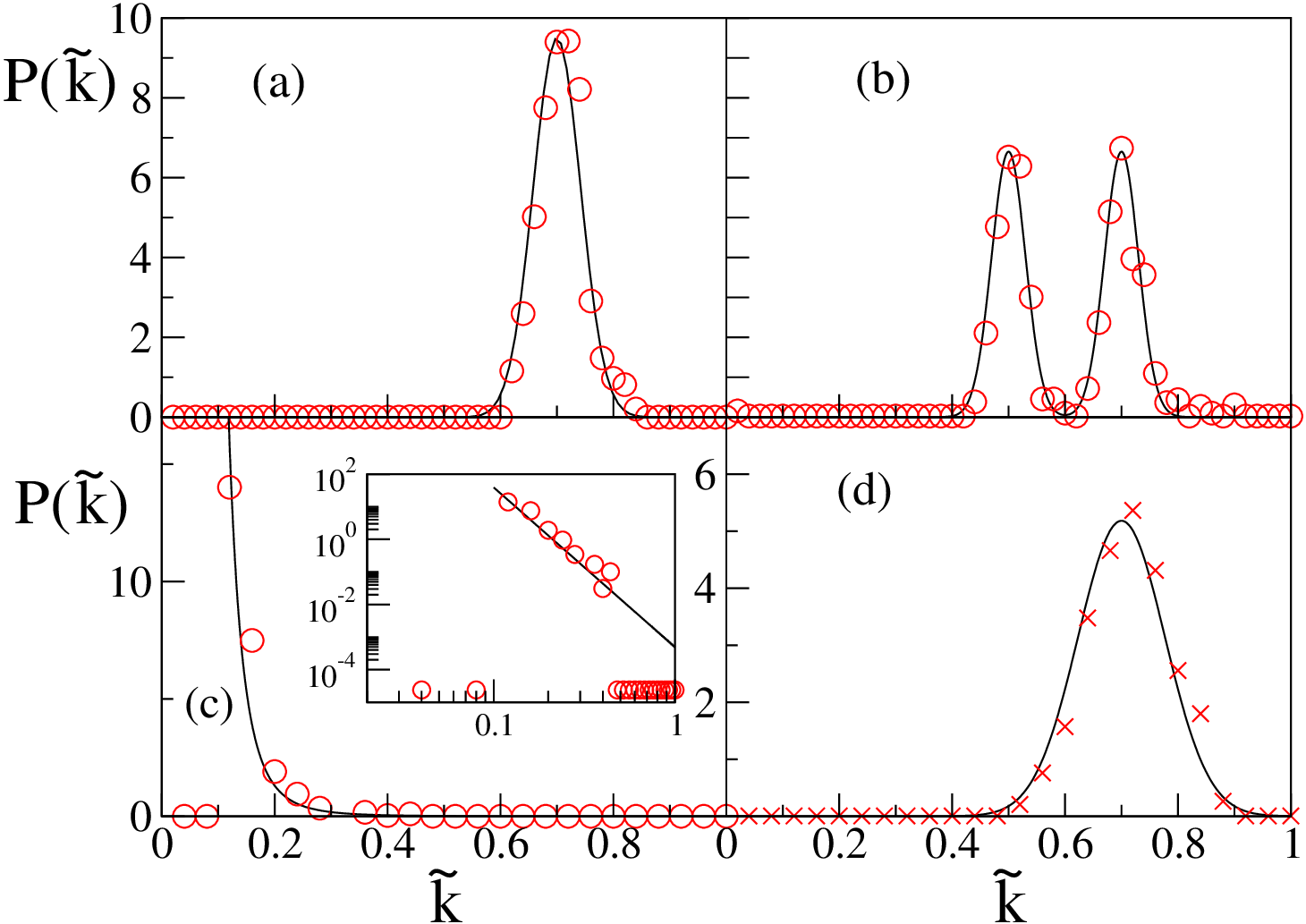}
\caption{Inverse problem for $P(\tilde k)$ from the global field $Y(t)$. 
Panels (a), (b) and (c)  
show  three distributions  of the kind considered in Fig. (4) (black continuous curves) 
for the HMF equations and their reconstructions (circles) by the inverse method.  The parameters of the three distributions are
$\sigma_{\tilde k}=0.043$, $\tilde k_2=0.7$ and $\alpha =4.9$.
In panel (d) we show the reconstruction (crosses) of  $P(\tilde k)$ (black continuous line)  
by the average field $Y(t)$ generated by the dynamics of a {\sl finite size} network with $N=500$.  }
\label{4}
\end{figure}   

To check our inverse method, we choose a distribution $P(\tilde k)$, evolve the system and extract the global synaptic field $Y(t)$.  We then verify if the procedure reconstructs correctly the original distribution $P(\tilde k)$. In  panels (a), (b) and (c) of Fig. \ref{4} we show examples in which $Y(t)$ has been obtained from the simulation of the HMF with different $P(\tilde k)$ (Gaussian, double peak Gaussian and power law).  We can see that the method determines confidently the original distribution $P(\tilde k)$. Notice that the method fails as soon as the locked component disappears, as explained in the methods section. Remarkably, the method can recognize
the discontinuity of the distribution in $\tilde k=\tilde k_{\mathrm{min}}$  and the value of the exponent of the power law $\alpha=4.9$.
Finally, in panel (d) of Fig.\ref{4}, we show the result of the inverse problem for the distribution $P(\tilde k)$ obtained from a global signal generated by a finite size realization with $N=500$ and $\langle k\rangle=350$. The significant agreement 
indicates that the HMF and its inverse problem are able to infer the in--degree probability distribution $P(\tilde k)$ even for a realistic finite size network. 
This last result is particularly important, as it opens new perspectives for experimental data analysis, where the average neural activity is typically measured from finite size samples with finite but large connectivity.

\section*{Discussion}

The direct and inverse problem for neural dynamics on random networks are both accessible through the HMF approach  proposed in this paper. The mean-field equations provide a semi--analytic form for the return map of the firing times of neurons and they allow to evidence the effects of the in-degree distribution on the synchronization transition. This phenomenology is not limited to the LIF model analyzed here and it is observed in several neural models on random networks. We expect that the HMF equations could shed light on the nature of this interesting, but still not well understood, class of synchronization transitions \cite{ps,olmi,luccioli}. The mean field nature of the approach does not represent a limitation, since  the equations are expected to give a faithful description of the dynamics also in networks with finite but large average in-degree, corresponding to the experimental situation observed in many cortical regions \cite{kgrande}.

The inverse approach, although tested here only on numerical experiments, gives excellent results on the reconstruction of a wide class of in-degree distributions and it could open new perspectives on data analysis, allowing to reconstruct the main topological features of the neural network producing the QSE. The inversion appears to be stable with respect to noise, as clearly shown by the effectiveness of the procedure tested on a finite realization, where the temporal signals of the average synaptic activity is noisy. We also expect our inverse approach to be robust with respect to the addition of limited randomness in leaky currents, and also with respect to a noise compatible with the signal detection from instruments. Finally, the method is very general and it can be applied to a wide class of dynamical models on networks, as those described in Eq. (\ref{eq1n}).

Further developments will allow to get closer to real experimental situations. We mention  the most important, i.e.  the introduction of inhibitory neurons and the extension of our approach by taking into account networks with degree correlations \cite{mendes},  that are known to characterize real structures, and sparse networks. The HMF equations appear to be sufficiently flexible and simple to allow for these extensions.  

\section*{Methods}

\subsection*{Montecarlo approach to the inverse problem}

In this section we provide details of the algorithmic procedure adopted for solving the
inverse problem, i.e. reconstructing the distribution $P(\tilde k)$ from Eq. (\ref{global}).
In the HMF formulation, the field $Y(t)$ is generated by an infinite number of neurons and $\tilde k$ is a continuous
variable in  the interval $ (0,1]$. In practice, we can sample uniformly this unit  interval by $L$ disjoint subintervals
of length $1/L$, labelled by the integer $i$. This corresponds to an {\sl effective neural index} $i$, that identifies the class
of neurons with in-degree $\tilde k_i = i/L$. In this way we obtain a discretized definition converging to  Eq.(\ref{global}) for $L \to \infty$:
\begin{equation}
Y(t)= \int_0^1  P(\tilde k) \mathsf{y}_{\tilde k}(t)d\tilde k 
      \simeq \frac{1}{L}\sum_{i=0}^{L-1} P(\tilde k_i) \mathsf{y}_{\tilde k_i}(t)  \,\, .
\end{equation} 
In order to improve the stability and the convergence of the algorithm 
by smoothing
the fluctuations of the fields $\mathsf{y}_{\tilde k_i}(t)$,  it  is convenient to consider a coarse--graining of the sampling by approximating 
$Y(t)$ as follows  
\begin{equation}
Y(t)=\frac{1}{L'} \sum_{i=0}^{L'-1}P(\tilde k_i)\langle \mathsf{y}_{\tilde k_i}(t)\rangle.
\label{discret_freholm}
\end{equation}
where $\langle \mathsf{y}_{\tilde k_i}(t)\rangle$ is the average of $L/L'$ synaptic fields of connectivity $\tilde k \in [{\tilde k_i},{\tilde k_{i+1}}]$.
This is the discretized Fredholm equation that one can solve to obtain $P(\tilde k_i)$ from the knowledge of $\langle \mathsf{y}_{\tilde k_i}(t)\rangle$ and $Y(t)$. 
For this aim we use a Monte Carlo (MC) minimization procedure, by introducing at each MC step, $n$, a trial solution, $P_n(\tilde k_i)$, in the form of a normalized non-negative in-degree
distribution.
Then, we evaluate the field $Y_n(t)$ and the distance $\gamma_n$ defined as:
\begin{align}\label{variance}
Y_n(t,P_n(\tilde k_i)) &=\frac{1}{L'} \sum_{i=0}^{L'-1}P_n(\tilde k_i)\langle \mathsf{y}_{\tilde k_i}(t)\rangle\\
 \gamma_n(P_n(\tilde k_i))^2 &=\frac{1}{t_2-t_1}\int_{t_1}^{t_2}  \frac{\Big[Y_n(t,P_n(\tilde k_i))-Y(t))\Big]^2}{Y^2(t)}dt \,\,\, .
\end{align}
The time interval $[t_1,t_2]$ has to be taken large enough to obtain
a reliable estimate of $\gamma_n$. For instance, in the case shown
in Fig.\ref{rp1}, where $Y(t)$ exhibits an almost periodic evolution of
period $T\approx 1$ in the adimensional units of the model, we have
used $t_2 - t_1 = 10$. The overall configuration of the synaptic fields, at iteration step $n+1$, is obtained by choosing randomly two 
values  $\tilde k_j$ and $\tilde k_l$, and by defining a new trial solution
$\bar{P}_{n+1}(\tilde k)=P_n(\tilde k)+\epsilon \delta_{\tilde k, \tilde k_j}-\epsilon \delta_{\tilde k, \tilde k_l}$, so that, provided both $\bar P_{n+1}(\tilde k_j)$ and
$\bar P_{n+1}(\tilde k_l)$ are non-negative, we increase and decrease $P_n(\tilde k_j)$ of the same amount,
$\epsilon$, in $\tilde k_j$ and $\tilde k_l$ respectively. A suitable choice is $\epsilon \sim \mathcal{O}(10^{-4})$.
Then, we  evaluate $\gamma_{n+1}(\bar{P}_{n+1}(\tilde k_i))$: 
If $\gamma_{n+1}(\bar{P}_{n+1}(\tilde k_i))<\gamma_n({P}_n(\tilde k_i))$
the step  is accepted i.e. $P_{n+1}=\bar P_{n+1}$, otherwise $P_{n+1}=P_n$. This MC procedure
amounts to the implementation of a {\sl zero temperature dynamics}, where the cost function $\gamma_n({P}_n(\tilde k_i))$ can only decrease. 
In principle, the inverse problem  in the form of  Eq.(\ref{discret_freholm}) is solved, i.e. $Y_n(t,P_n(\tilde k_i))=Y(t)$,
if $\gamma_n({P}_n(\tilde k_i))=0$. In practice, the  approximations introduced by the coarse-graining procedure
do not allow for a fast convergence to the exact solution, but $P_n(\tilde k_i)$ can be considered a reliable reconstruction
of the actual $P(\tilde k)$ already for $\gamma_n < 10^{-2}$. 
We have checked that the results of the MC procedure are quite stable with respect to different choices of the initial conditions ${P}_0(\tilde k_i)$, thus confirming  the robustness of the method. 
 We give in conclusion some comments on the very definition of the coarse-grained synaptic field
 $\langle \mathsf{y}_{\tilde k_i}(t)\rangle$.
Since small differences in the values of $\tilde k_i$ reflect in small differences in the dynamics, for not too large intervals $[{\tilde k_i},{\tilde k_{i+1}}]$ the quantity  $\langle \mathsf{ y}_{\tilde k_i}(t)\rangle$ can be considered as an average over different initial conditions. For locked neurons the convergence of the average procedure defining $\langle \mathsf{ y}_{\tilde k_i}(t)\rangle$ is quite fast, since all the initial conditions  tend to the stable fixed point, identified by the
return map described in the previous subsection. On the other hand,  the convergence of the same quantity  
for unlocked neurons  should require an average over a huge number of initial conditions. 
For this reason, the broader is the distribution, i.e. the bigger is
the unlocked component (see Fig.\ref{cross}), the more
computationally expensive is the solution of the inverse problem. This
numerical drawback for broad distributions emerges in our tests of the
inversion procedure described in Fig. \ref{4}. Moreover, such tests show that the procedure works
insofar the QSE are not negligible, but it fails in the absence of the locking mechanism. 
In this case, indeed, the global field $Y(t)$ is constant and also
$\langle \mathsf{ y}_{\tilde k_i}(t)\rangle$ become constant,
 when averaging over a sufficiently large number of samples. 
This situation makes Eq.(\ref{discret_freholm}) trivial and useless to evaluate $P(\tilde k_i)$.
We want to observe that, while in general $\mathsf{ y}_{\tilde k_i}(t) \not = y_{\tilde k_i}(t)$, one can
reasonably expect that  $\langle \mathsf{ y}_{\tilde k_i}(t)\rangle$ is a very good approximation
of $\langle  y_{\tilde k_i}(t)\rangle$. This remark points out the conceptual importance of the HMF formulation
for the possibility of solving  the inverse problem.

\section*{Acknowledgements}

This work is partially supported by  the Centro Interdipartimentale per lo
Studio delle Dinamiche Complesse (CSDC) of the Universita' di
Firenze, and by the Istituto Nazionale di Fisica Nucleare (INFN).

\section*{Additional Information}

\subsection*{Author contribution statement}
R. B., M. C., M. d. V., R. L., A, V.  contributed to the formulation of the problem, to its solution,
to the discussions, and to the writing of the paper.
M. d. V. performed the simulations and produced all the plots.

\subsection*{Competing financial interests}
The authors declare no competing financial interests.

\newpage

\section{Captions}

Fig. \ref{rp1}:  Raster plot representation of the dynamics of a network of $N=500$ neurons with a Gaussian in-degree distribution $P(\tilde k)$
($\langle \tilde k \rangle=0.7$,  $\sigma_{\tilde k}=0.077$). Black dots signal single firing events of neurons at time $t$.
Neurons are naturally ordered along the vertical axis according to the values of their in-degree. 
The  global field $Y(t)$ (red line) is superposed to the raster plot for comparison (its actual
values are multiplied by the factor $10^3$, to make it visible on the vertical
scale).  \\

Fig. \ref{2}:  Time average of inter-spike intervals $\overline{ISI}({\tilde k})$ vs $\tilde k$ from a Gaussian distribution  with  $\langle \tilde k \rangle=0.7$ and $\sigma_{\tilde k}=0.077$
and for three networks with $N=500$ (blue triangles), $N=5000$ (red diamonds), $N=20000$ (green squares). For each size, the average is taken over 8 different realizations of the random network.  We have also performed a suitable binning over the values of $\tilde k$, thus yielding the numerical 
estimates of the critical values $\tilde k_{c1} \approx 0.49$ and $\tilde k_{c2} \approx 0.70$.
 In the inset we show a zoom of the crossover region  close to $\langle \tilde k \rangle =0.7$. Black dots are the result of simulations of the mean field dynamics
(see Eq.s  (6) -- (8)) with $M=307$. \\

Fig. \ref{6}: The return map $R_{\tilde k}$ in Eq. (10) of the rescaled variables $t_{\tilde k}(n)/T$ for different values of $\tilde k$, corresponding to lines of different colors, according to the legend in the inset: the black line is the bisector of the square.\\

Fig. \ref{cross}:  The fraction of locked neurons, $f_l$, as a function of the standard deviation $\sigma_{\tilde k}$ of the distributions: truncated 
Gaussian  with $\langle \tilde k \rangle  = 0.7$ (black dots);  truncated superposition of 
two Gaussians (both with standard deviation 0.03), one centered at  $\tilde k_1=0.5$ and the other one at a varying value $\tilde k_2$, that determines the
overall standard deviation $\sigma_{\tilde k}$ (blue squares); truncated power law distribution  with $\tilde k_{\mathrm{min}}=0.1$
(red diamonds). In the last case the value of the standard deviation is changed by varying the exponent $\alpha$, while the average $\langle \tilde k\rangle$ changes accordingly.  Lines have been drawn to guide the eyes.\\

Fig. \ref{4}: Inverse problem for $P(\tilde k)$ from the global field $Y(t)$. 
Panels (a), (b) and (c)  
show  three distributions  of the kind considered in Fig. (4) (black continuous curves) 
for the HMF equations and their reconstructions (circles) by the inverse method.  The parameters of the three distributions are
$\sigma_{\tilde k}=0.043$, $\tilde k_2=0.7$ and $\alpha =4.9$.
In panel (d) we show the reconstruction (crosses) of  $P(\tilde k)$ (black continuous line)  
by the average field $Y(t)$ generated by the dynamics of a {\sl finite size} network with $N=500$.

\end{document}